\documentclass[10pt,a4paper]{article}

\usepackage{epsfig}

\usepackage{graphicx}
\usepackage{amsmath}
\usepackage{amssymb}
\usepackage{psfrag}
\usepackage[sans]{dsfont}

\DeclareMathOperator{\Entropy}{S}

\newcommand{\capacity}{C^*=\lim_{n\rightarrow \infty}\frac{1}{n}C^1_{\textrm{Classical}}(\Lambda_n))= 1- \lim_{n\rightarrow \infty}\frac{1}{n} \Entropy(\Lambda_n(\rho))}
\newcommand{\entropy}{s(\mu) = \sum_{a \in L} \int_{\mathcal{B}} \phi(d\nu) h_a (\nu)}

\begin{document}

\title{The Algebraic Measure of a Hidden Markov Quantum Memory Channel}
\author{I. Akhalwaya$^1$, J. Wouters$^2$, M. Fannes$^2$, F. Petruccione$^1$\\
\small
$^1$Quantum Research Group and National Institute for Theoretical Physics\\ \small 
School of Physics, UKZN, Private Bag X54001, Durban, 4000, RSA\\ \small
$^2$ Instituut voor Theoretische Fysica, Katholieke Universiteit Leuven, \\ \small
Celestijnenlaan 200D, B-3001 Heverlee, Belgium
}

\date{21 September 2008}






\maketitle

\begin{abstract}
This paper was presented in poster form at and in the proceedings of the QCMC 2008. It is a summary of a fuller paper to appear separately. The classical product state capacity of a noisy quantum channel with memory is investigated. A forgetful noise-memory channel is constructed by Markov switching between two depolarizing channels which introduces non-Markovian noise correlations between successive channel uses. This function of a Markov process can be reformulated as an algebraic measure. This framework provides an expression for the asymptotic entropy rate and thus enables the calculation of the classical capacity. The effects of the hidden-Markovian memory on the capacity are explored. An increase in noise-correlations is found to increase the capacity.
\end{abstract}

\section{Introduction}

Great strides have been made in understanding the capacity of quantum channels. For example, the celebrated Holevo-Schumacher-Westmoreland (HSW) theorem \cite{HSW} gives an expression for the classical capacity of a noisy quantum channel with product state inputs. The memoryless channel restriction has since been extended to, so called, \textit{forgetful} memory channels \cite{DKRW}. The inclusion of memory is the next step in the attempt of accurately modelling the complicated noise-correlated real world. 

We construct a forgetful channel and incorporate memory effects by Markov switching between two sub-channels. We investigate the classical product state capacity of this channel. This involves the entropy of the classical output of the channel. To manage this complicated conditional dependence, we use the hidden Markov nature of the process to reformulate the problem using the algebraic measure construction \cite{FNS91}. The algebraic measure approach allows us to derive an expression for the asymptotic entropy rate. We then explore the effects that our non-Markovian memory has on the classical product state capacity.

\section{Construction of the Channel}
\label{sec:construction}

The forgetful channel is constructed by combining two memoryless single qubit depolarizing channels ($ \mathcal{E}_0$ and $ \mathcal{E}_1$), switching between them using a two-state Markov chain ($Q=(q_{ij}), \quad i,j \in \lbrace 0,1\rbrace$). Thus, $Q$ is the $2 \times 2$ Markovian channel selection matrix with $q_{ij}$ being the probability of switching from channel $i$ to channel $j$.

The depolarizing channels can be written as: $ \mathcal{E}_i(\rho) = x_i^0 \rho + x_i^1 (\mathbf{1} - \rho)$. These single qubit channels can be thought of as probabilistically mixing the identity channel (with probability $x_i^0$) and `flip' channel (with probability $x_i^1=1-x_i^0$) acting on a single qubit density operator $\rho$. However this rewriting is only completely positive for $1/3 \leq x^0_i \leq 1$. 

The built-up channel corresponding to $n$ successive uses of the single qubit sub-channels, $\Lambda_n$, is constructed as follows:
\[\Lambda_n = \rho_1\otimes\ldots\otimes\rho_n \mapsto \sum_{i_1,\ldots,i_n} \gamma_{i_1} q_{i_1i_2} \ldots q_{i_{n-1}i_n}  \mathcal{E}_{i_1}(\rho_1)\otimes\ldots\otimes \mathcal{E}_{i_n}(\rho_n) \; .\]
The sum is over all possible Markov paths $(i_1,\ldots,i_n)\in \lbrace 0,1\rbrace^n$ and each term is a tensor product of the selected sub-channels weighted by the probability of occurrence ($\gamma_i$ is the initial probability of selection). We calculate the capacity with this $n$-use form of the channel and then regularise to take the limit as $n$ goes to infinity. The HSW theorem \cite{HSW} built on by the forgetful channel extension \cite{DKRW} gives us an expression for the capacity $C^1_{\textrm{Classical}}$ (classical product state):
\[C^1_{\textrm{Classical}}(\Lambda_n)= \chi^*(\Lambda_n)\;,\]
where $\chi^*$ is the maximisation over all possible input ensembles (effectively only pure states) of the Holevo  $\chi$ quantity.

For this channel, the maximum is obtained using the uniformly distributed computational basis states \cite{KIN03}. By taking the asymptotic average of the $n$-product classical capacity, we arrive at the capacity of the memory channel \cite{DKRW}:
\begin{equation}
\capacity
\; .
\end{equation}

\section{Algebraic Measure of the Channel}
\label{sec:algebraic}

%
An algebraic measure, $\mu$, is a translational-invariant measure on a set $\lbrace 0, \ldots, q-1\rbrace^\mathbb{Z}$, with probabilities determined by matrices $E_a$ with positive entries, one for each of the $q$ states. The probability of a sequence is obtained by applying a positive linear functional $\sigma$ to a matrix product of the corresponding matrices of the states of the sequence: $\mu(i_1,\ldots,i_n)=\sigma(E_{i_1}\ldots E_{i_n})$. This matrix algebraic construction is the reason for the name \textit{Algebraic Measure}, studied in detail in Ref. \cite{FNS91}. Every measure that is a function of a Markov process (also called hidden Markov) is an algebraic measure and remarkably, the converse holds too. The relationship between the hidden Markov measure, say $\mu'$ on $K^{\mathbb{Z}}$, and the underlying Markov measure $\nu$ with the Markov property on $L^{\mathbb{Z}}$ is through a `tracing' function $\Phi : L \rightarrow K$, as follows ($L$ and $K$ are finite sets of states):
\[ \mu'((\omega_m, \ldots, \omega_n)) = \sum_{\substack{\epsilon_m,\ldots,\epsilon_n\\ \Phi(\epsilon_m)=\omega_m \ldots \Phi(\epsilon_n)=\omega_n}} \nu((\epsilon_m,\ldots,\epsilon_n)) \; ,\]
where $\omega_m, \ldots, \omega_n \in K$ and $\epsilon_m,\ldots,\epsilon_n \in L$.



The underlying Markov process for the overall quantum channel has a four state configuration space corresponding to channel selection and error occurrence: $K = \lbrace (00), (01), (10), (11) \rbrace$. The first index indicates which depolarizing channel has been chosen and the second indicates whether a bit flip occurred. The transition matrix, $E$, for this process is then given by $E_{(ij)(i'j')} = q_{ii'}x_{i'}^{j'}$: the probability of going from $(ij)$ to $(i'j')$ is given by the switching probability from channel $i$ to $i'$, $q_{ii'}$, multiplied by the probability that channel $i'$ produces the error-configuration $j'$, $x_{i'}^{j'}$.

The function that produces the correct hidden Markov process is then given by
\[ \Phi((i,j))=j \; .\]
This function reflects the fact that we are unaware of the choice of channel that has been made. The only effect that is visible from the outside is whether or not an input qubit has been flipped. Thus, $\Phi$ has to `trace out' the choice of channel. $\Phi$ maps into the two-state error configuration space containing `no flip' and `flip':  $L = \lbrace 0,1 \rbrace$ .

With the successful construction of the algebraic measure in terms of the traced out matrices, we try to calculate the asymptotic entropy rate. In \cite{FNS91,DB57}, an expression for this rate was derived:
\begin{equation}
\label{eq:entropy}
\entropy
\; . \end{equation}
where $\mathcal{B}$ is the set of probability measures on the set of $q$ states and $h_a$ is an entropy-like function. The measure $\phi$ is the fixed point of a transformation $T_\mu$ determined by the matrices of the algebraic measure $E_a$. We have used this transformation to generate $\phi$ iteratively. The iterative procedure can be simulated numerically allowing us to approximate the entropy and thus the capacity.

\section{Results}



In order to simplify interpretation, we symmetrise the Markov switching process by the substitution $q_{00} \rightarrow(1 + s)/2$, $q_{10} \rightarrow(1 - s)/2$. We parametrize the error probabilities by their average and difference: $x_0^0 \rightarrow a + d$, $x_1^0 \rightarrow a - d$. By defining $a$ as the average probability of no-error of the two sub-channels and studying the difference, $d$, we are pre-empting the result that it is the average and difference of the two sub-channels that are important to the capacity and not the absolute probabilities.

The main result is that the capacity increases with stronger noise-correlations which manifests itself in two ways. Firstly, if we make the switching more correlated ($s$ away from $0$) the capacity increases and secondly, if we increase the difference between the two sub-channels the capacity also increases.

\begin{figure}
\centering
{
\footnotesize

\psfrag{x}[Tl][br]{$a$}
\psfrag{y}[Tr][bl]{$s$}
\psfrag{z}[cl][cr]{$C^*$}

\psfrag{a1}[Tl][Tl]{$0.3$}
\psfrag{a2}[Tl][Tl]{$0.6$}
\psfrag{a3}[Tl][Tl]{$1$}

\psfrag{b1}[Tr][Tr]{$0$}
\psfrag{b2}[Tr][Tr]{$0.5$}
\psfrag{b3}[Tr][Tr]{$1$}

\psfrag{c1}[bl][cl]{$0$}
\psfrag{c2}[bl][cl]{}
\psfrag{c3}[bl][cl]{$0.7$}

\includegraphics[width=0.9 \textwidth]{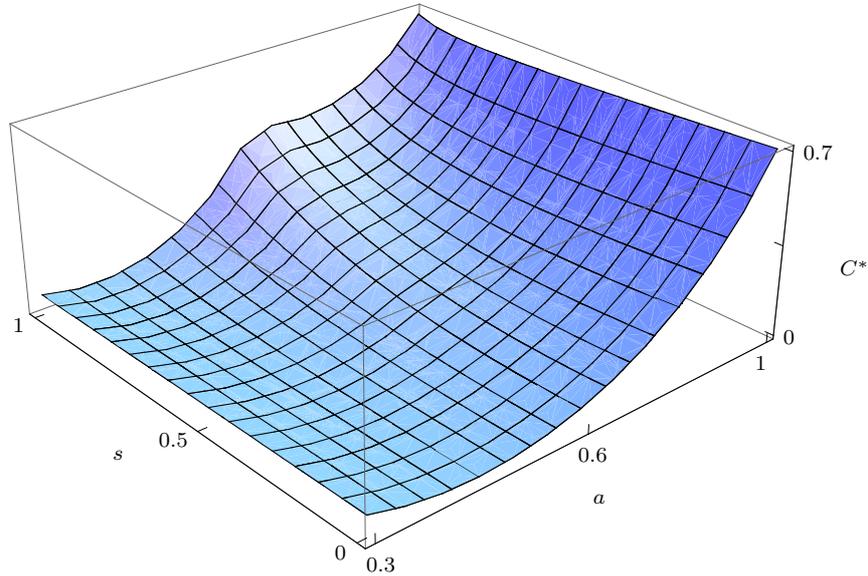}
}
\caption{Noisy-on-average but Distinct allows increase in Capacity}
\label{fig:capacity}
\end{figure}

In Figure \ref{fig:capacity}, $d$ is set to the maximum possible value while keeping an average of $a$ ($d=\min[a-1/3,1-a]$). The capacity is plotted against varying $a$ and $s$. We can see that the capacity increases as the noise-correlation ($s$) gets stronger. When $a=2/3$, $d$ attains its maximum ($1/3$) and the effect of increasing $s$ on the capacity is greatest. Another interesting observation is the case when the two sub-channels average to the maximally mixing channel ($a=1/2$, ordinarily with zero capacity), taking into account memory effects there is a non-zero capacity.

\section{Conclusion}
We have constructed a simple forgetful noise-memory quantum channel. The noise-correlation is a function of the underlying hidden Markov process which allowed us to construct the algebraic measure. We used the measure in an algebraic asymptotic entropy expression. Without this, the entropy would be very difficult to compute, involving exponentially many Markov paths.

We studied the effects that the noise correlations had on the classical capacity and discovered that the capacity increases with stronger correlations. This is sensible because the correlations can be used to combat the noise when coding information.

\section{Acknowledgments}
We would like to acknowledge N. Datta and T. Dorlas for the idea of the channel construction and valuable assitance. This work is based upon research supported by the South African Research Chair Initiative of the Department of Science and Technology and National Reseach Foundation.

\end{document}